# Are gravitational constant measurement discrepancies linked to galaxy rotation curves ?

Norbert Klein, Imperial College London, Department of Materials, London SW7 2AZ, UK

The discrepancies between recently reported experimental values of the gravitational constant were analysed within an inertia interpretation of MOND theory. According to this scenario the relative gravitational acceleration between a test mass and an array of source masses determines the magnitude of post Newtonian corrections at small magnitudes of acceleration. The analysis was applied to one of the most advanced recent Cavendish-type experiment which revealed an experimental value for the gravitational constant of 180 ppm above the current CODATA value with more than five standard deviations significance. A remarkable agreement between this discrepancy and the acceleration anomalies inherent of galaxy rotation curves was found by a consistent extrapolation within the framework of MOND. This surprising result suggests that the two anomalies on totally different length scales may originate from the same underlying physics.

More than 300 years after Newton, gravitation still remains one of the great miracles, and its understanding in the framework of a unified theory with the other fundamental forces is still lacking [1,2]. Newton's gravitational law and its subtle modifications within the framework of General Relativity allow highly accurate predictions of the motion of planets and other objects within our solar system. However, as first reported by Vera Rubin in 1970, the measured rotation of galaxies is incompatible with Newton's law, if only visible matter is considered for modelling the dynamic behaviour [3]. Therefore, the concept of dark matter was suggested as a possible explanation of this discrepancy, but dark matter has not yet been observed directly [4]. As an alternative to dark matter, the concept of Modified Newtonian Dynamics (MOND) was introduced by Milgrom in 1983 [5], which accounts for deviations from Newton's law in case of small acceleration magnitudes. The universal radial acceleration relation in rotationally supported galaxies, as presented recently by Mc Gaugh et al. [6], cannot be consistently explained by dark matter, but strongly favours the validity of MOND theory [7].Although MOND is still a phenomenological theory without any solid underpinning by physical principles, Milgrom pointed out recently that the underlying physics of MOND effects may be due to a modification of inertia, as a result of the interaction of accelerated matter with the quantum vacuum [8]. In contrast to MOND interpretations which suggest a modification of the gravitational force law, as being formulated by Bekenstein and Milgrom through a modified Poisson equation for the gravitational potential [9], the inertia interpretation would allow the observation of MOND effects in earthbound laboratories. One requirement for this is a sufficient isolation of the relevant test masses from seismic noise, which ensures that seismic accelerations (and others) are smaller than the measured acceleration due to the gravitational force to be measured (see discussion).

Accurate measurements of the gravitational constant $G$, most of them by Cavendish-type torsion balances [1,10,11], Fabry - Perot microwave resonators (direct involvement of the author) [12,13] and optical interferometers [14] suspended as linear pendulums are not in perfect agreement with each other, and the observed discrepancies of up to ten standard deviations remain unexplained to date [1]. Noticeably, the recently reported results by Quinn and Speake [10], as determined by two independent operation methods of a Cavendish-type $G$ torsion balance, are 180 ppm above the results obtained by Schlamminger et al. [15], which are based on a beam balance employing 13 tons of liquid mercury as source mass – the latter being consistent with the most recent CODATA value [16]. As a possible explanation for these discrepancies, Anderson et al. suggested a sinusoidal time variation of *G* according to an analysis of *G* experiments being performed since year 1980 [17], but according to a recent analysis of the consequence for orbital motions within our solar system by Iorio this scenario is in contradiction to the experimental constraints for the observed orbit increase of the LAGEOS satellite and anomalous perihelion precession of Saturn [18].

Inspired by the analysis of the acceleration magnitude dependence of gravitational force measurements by a pendulum gravimeter within MOND theory by Meyer et al. [13], in this report a post-Newtonian analysis of selected *G* measurements is pursued and compared to galaxy rotation curves using different MOND extrapolation functions. In order to adapt MOND theory for *G* experiments, two point masses $m_1$ and $m_2$ being separated by a distance $r_{12}$ are considered. In an idealized *G* experiment, one of the two masses (called "test mass") is moved by an incremental distance from one equilibrium position (source mass at infinite distance from the test mass) to another (source mass at distance $r_{12}$ from the test mass) due to a small acceleration caused by the gravitational force $\vec{F}_{12}$ between $m_1$ and $m_2$. In a centre-of-mass frame of reference, the Newtonian dynamics is described in relative coordinates, the inertia in Newton's second law is determined by the reduced mass $\mu$.

$$\vec{F}_{12N} = \mu \vec{a}_{rN} \qquad \vec{a}_{rN} = \vec{a}_{1N} - \vec{a}_{2N} \qquad \mu = \frac{m_1 m_2}{m_1 + m_2} \qquad \text{Eq. 1}$$

Although the two masses are arranged within a non-inertial frame of reference (the earth), it is presumed here that the post-Newtonian effect is solely determined by the inertia of the reduced mass with respect to the gravitational attraction between $m_1$ and $m_2$. A torsion pendulum represent a suitable approximation for a local inertial-frame of reference with respect to the direction of test mass motion caused by the gravitational force between test- and source masses. In order to fit MOND-type post-Newtonian corrections at low magnitudes of acceleration, a universal correction function $f(|\vec{a}_{rN}|)$

$$\vec{a}_r = \vec{a}_{rN} f(|\vec{a}_{rN}|); \qquad f(|\vec{a}_{rN}|) = \left[1 + \left(\frac{a_0}{|\vec{a}_{rN}|}\right)^\alpha\right]^{\frac{1}{2\alpha}} \qquad \text{Eq. 2}$$

is suggested, with $a_0 = 1.2 \cdot 10^{-10}$ m/s² resembling the fundamental MOND acceleration parameter [5,7,8] and $\alpha$ is a free parameter. The function $f(|\vec{a}_{rN}|)$ represents a correction of the relative Newtonian acceleration $\vec{a}_{rN}$ and becomes significant for small magnitudes of the latter. For the case of a negligible test mass ($m_1 \ll m_2$, $\mu = m_2$, $a_{rN} = a_N$) Eq. 2 fulfils the requirements for the asymptotic behaviour according to MOND theory, $a(a_N \gg a_0) \to a_N$ (Newtonian limit) and $a(a_N \ll a_0) \to (a_0 a_N)^{1/2}$ (deep MOND limit) [5,7,8]. The latter explains the Tully-Fisher relation [19], one of the hallmarks of MOND theory. The recent discovery of Dragonfly 44 [20], a galaxy of the size of the Milky Way but with only 1% of its visible mass, celebrated in the press as galaxy composed of 99.99% dark matter, may be considered as new hallmark of MOND theory, because the measured velocity dispersion of this galaxy agrees with the predicted value for the deep MOND limit, which is independent of the choice of the extrapolation function [21].

Within a modified inertia interpretation of MOND, the correction of the relative acceleration is equivalent to a modification of the reduced mass according to Eqs. 1 and 2.

$$\mu_{MOND} = \frac{m_1 m_2}{m_1 + m_2} f^{-1}(|\vec{a}_{rN}|); \qquad \text{Eq. 3}$$

The correction function according to Eq. 2 will be compared with extrapolation functions representing solutions of the modified Poisson equation [9] for one point mass for the most commonly used MOND extrapolation functions $\mu_{MOND0}(x)=[(1+4x)^{1/2}+]/[(1+4x)^{1/2}-1]$, $\mu_{MOND1}(x)=x/(1+x)$ and $\mu_{MOND2}(x)=x/(1+x^2)^{1/2}$ [8,13], which are

$$a_{MOND0} = a_N \left[ 1 + \sqrt{\frac{a_0}{a_N}} \right] \qquad \text{Eq. 4}$$

$$a_{MOND1} = a_N \left[ \frac{1}{2} + \sqrt{\frac{1}{4} + \frac{a_0}{a_N}} \right] \qquad \text{Eq. 5}$$

$$a_{MOND2} = a_N \sqrt{\frac{1}{2} + \frac{1}{2}\sqrt{1 + \left(\frac{2a_0}{a_N}\right)^2}} \qquad . \qquad \text{Eq. 6}$$

It is worth noting that $a_{MOND0}$, which results from the relativistic generalization of MOND theory (TeVeS) [22], is identical to Eq. 2 for the parameter choice of $\alpha = \frac{1}{2}$. The practical advantage of the suggested extrapolation function (Eq.2) is that the smoothness of the transition from the Newtonian regime to the deep MOND regime can be varied continuously by the choice of the parameter $\alpha$. The inertia interpretation of MOND does not constrain the extrapolation functions to simple solutions of the modified Poison equation, therefore Eq. 2 is not in contradiction to any of the fundamental MOND tenets.

Inserting Newton's law of gravity into Eq. 2 and using the definition for the reduced mass (Eq. 1), Eq. 2 turns out to be mathematically equivalent to a modified gravitational force law

$$\vec{F}_{12} = G m_1 m_2 \frac{\vec{r}_{12}}{|\vec{r}_{12}|^3} \left[ 1 + \left( \frac{a_0 |\vec{r}_{12}|^2}{G(m_1+m_2)} \right)^\alpha \right]^{\frac{1}{2\alpha}} \qquad . \qquad \text{Eq. 7}$$

$F_{12}$ is symmetric in $m_1$ and $m_2$, i.e. Newton's 1st law is inherently fulfilled. Eq. 7 indicates that the MOND-corrected gravitational force exhibits a non-linear dependence on the masses $m_1$ and $m_2$ by which the force is generated. Due to the non-linear nature of the modified gravitational force, the superposition principle is violated, i.e. the effective force on a test mass $m_1$ by an ensemble of source masses $m_i$, $i = 2,…,N$ cannot be calculated by a simple vector superposition of forces according to Eq. 7.

As next step to analyse torsion balance experiments within MOND, the post-Newtonian correction for an array of source masses is evaluated. We consider $k$-1 point-type source masses $m_2….m_k$, located at positions $r_i$, $i=2..k$, and the test mass $m_1$ located at $r_1$. Usually, in a *G* experiment, only one component of this force is measured, determined by a unit vector ***n***. In a Cavendish-type *G* experiment, this component generates the measured torque.

The ***n***-component of the Newtonian acceleration of $m_1$ caused by the gravitational force between the array ($m_2...m_k$) and $m_1$ is given by

$$a_{1N} = G \sum_{i=2}^{k} \frac{m_i}{r_{1i}^2} \cos \varphi_i \qquad \text{Eq. 8}$$

with $\varphi_i$ representing the angle between $r_{1i}=r_i-r_1$ and ***n***, and $r_{1i}$ the distance between $m_1$ to $m_i$. Since the source masses $m_2,...,m_k$ are rigidly interconnected, it is appropriate to calculate the ***n***-component of acceleration of the entire array $m_2,...,m_k$ due to the gravitational force between the array and $m_1$.

$$a_{2..kN} = -\frac{G}{\sum_{i=2}^{k} m_i} \sum_{i=2}^{k} \frac{m_1 m_i}{r_{1i}^2} \cos \varphi_i \qquad \text{Eq. 9}$$

According to Eqs. 1, 8 and 9 the post-Newtonian correction is determined by the relative acceleration

$$a_{rN} = a_{2..kN} - a_{1N} = G \left( 1 + \frac{m_1}{\sum_{i=2}^{k} m_i} \right) \sum_{i=2}^{k} \frac{m_i}{r_{1i}^2} \cos \varphi_i \ . \qquad \text{Eq. 10}$$

The term in parenthesis deviates from unity for the case that the value of the test mass ($m_1$) cannot be neglected in comparison to the source masses ($m_2,...,m_k$). Hence, the post-Newtonian correction is determined by Eq. 2, employing Eq. 10 for the evaluation of the Newtonian relative acceleration.

Based on the methodology being described, Cavendish type experiments can be analysed for possible MOND effects. As outstanding recent experiments, the work by Schlamminger et al. (referred as $G_0$) [15] and Quinn and Speake [10] represent considerable steps forward in terms of the achieved accuracy and reproducibility. Noticeably, their results differ significantly from each other by 180 ppm – corresponding to five standard deviations of the experimental uncertainty.

$$G_0 = 6.674\,252(122) \cdot 10^{-11} \frac{\text{m}^3}{\text{kgs}^2} \qquad G_{Quinn} = 6.67545(18) \cdot 10^{-11} \frac{\text{m}^3}{\text{kgs}^2}$$

Schlamminger et al. used mercury source masses of 13,000 kg. Therefore, according to the criterion discussed before, this particular experiment is well within the deep Newtonian limit - within the claimed measurement error. We consider $G_0$ as a reference value for the gravitational constant, which is supported by the fact that $G_0$ lies well within the error limit of the current CODATA value of $(6.67384\pm0.0008)\cdot 10^{-11}$ m$^3$/kgs$^2$ [16]. Apparently, the value measured by Quinn and Speake is significantly higher.

In order to analyse the data within the suggested model, the geometry of the experiment by Quinn and Speake needs be analysed carefully. In Quinn's experiment the tangential component of the gravitational force between four field masses of $m_i$ = 11 kg ($i$=2,3,4,5) each, which are arranged on a circle of radius $R_2$ = 214 mm, and four test masses of $m_1$ = 1.2 kg each, arranged on a smaller concentric circle of radius of $R_1$ = 120 mm is determined. The circle of field masses is rotated by an angle $\varphi_0$ = 18.9° with respect to the circle of test masses, in order to maximize the torque. Although the field masses are short cylinders in Quinn's experiment, the point mass approximation is valid within an accuracy of a few percent, which is sufficient for the discussion of post-Newtonian corrections.

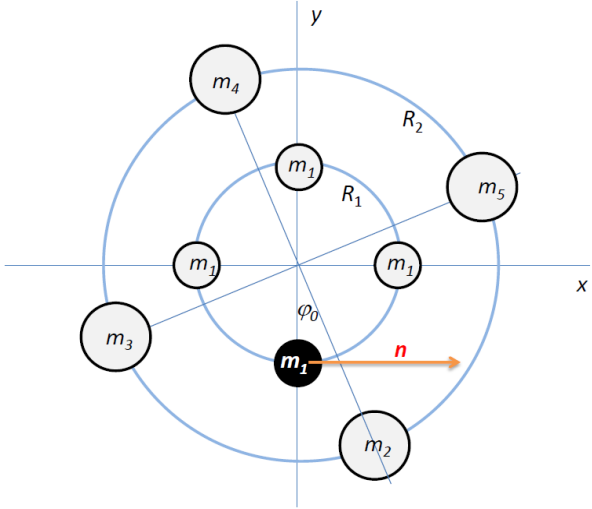

Fig. 1: Schematics of the Cavendish-type *G* experiment by Quinn et al. according to [10].

Using the geometry depicted in Fig. 1, the tangential component of the added gravitational acceleration of each field mass on one test is calculated from the geometry of the experiment. Due to the given symmetry, the force tangential component is the same for each test mass. The experiment is performed by measuring the torque generated by the gravitational force due to a rotation of the field masses from +$\varphi_0$ to -$\varphi_0$. According to Eq.10, the relative acceleration between one test mass and the array of four field masses is determined by working out the physical distance $r_{1i}$ and the angles $\varphi_i$ for each test mass from the geometry of the experiment depicted in Fig. 1. The resulting value of $a_{rN}$ is $6.14 \cdot 10^{-8}$ ms$^{-2}$ = 512·$a_0$.

In order to fit the $\alpha$ value of the extrapolation function to this result we employ

$$\frac{a_r}{a_{rN}} - 1 = \frac{G_{Quinn}}{G_0} - 1 = (1.93 \pm 0.45) \cdot 10^{-4} = \left[1 + \left(\frac{a_0}{a_{rN}}\right)^\alpha\right]^{\frac{1}{2\alpha}} - 1 \qquad \text{Eq. 11}$$

to account for possible post-Newtonian corrections. Based on the experimental value $G_0$ and $G_{Quinn}$ and their uncertainties, $\alpha$ ranges between 1.20 and 1.26, the best fit is achieved for $\alpha$=1.23. Obviously the fitted value of $\alpha$ depends on the choice of $a_0$.

In Fig. 2, the relative deviation between Quinn's and Schlamminger's result for *G* is presented as a function of the relative differential Newtonian acceleration being employed in Quinn's Cavendish experiment, the latter normalized to the MOND acceleration $a_0=1.2 \cdot 10^{-10}$ m/s$^2$. It is assumed that Schlamminger's result represents Newton's law. The area between the red lines corresponds to the extrapolated range of $\alpha$ values which are compatible with the analysis within the experimental error bars. For comparison, Eq. 2 is also plotted for a range of distinct values of $\alpha$ (full lines), along with MOND extrapolation functions according to Eqs. 3-5 (dashed lines).

As a direct comparison to recent astrophysical data, the purple dots represent individual resolved measurements along the rotation curves of nearly 100 spiral galaxies. The original data in [23] are presented as ratio of the squares of the measured orbital velocity and the calculated "baryonic" velocity, as being calculated from the Newtonian gravitational acceleration by the baryonic (= visible stars and interstellar gas) mass of the galaxy. Since the centrifugal acceleration is proportional to gravitational acceleration, this ratio is equal to the ratio of measured acceleration and Newtonian acceleration. This enables a direct comparison with the *G* data.

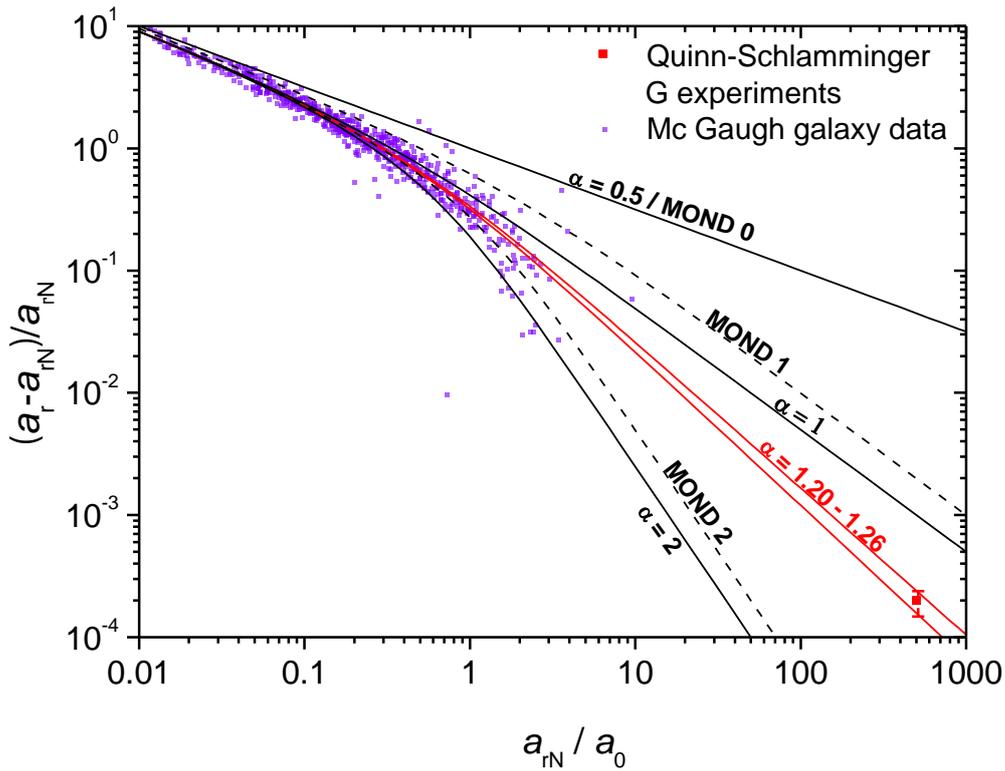

Fig.2: Comparison of the post-Newtonian acceleration determined from the Quinn-Schlamminger *G* discrepancy with galaxy rotation data, in terms of relative deviation from Newton's law, ($a_r/a_{rN}$)-1, as a function of the magnitude of the relative Newtonian acceleration $a_{rN}/a_0$. The experimental results are compared with a range of extrapolation functions according to MOND theories (see explanation in text). The red data point corresponds to the Quinn-Schlamminger *G* discrepancy, the purple data points represent the galaxy data, and the area between the red curves represents the extrapolation function based on the suggested post-Newtonian model for the range of the parameter $\alpha$ being compatible with the Quinn-Schlamminger discrepancy within their error margins.

The extrapolation function extracted from the Quinn-Schlamminger discrepancy gives an excellent fit to the cloud of galaxy rotation curve data and lies between MOND 1 (sometimes called MOND "simple" and MOND 2 (sometimes called MOND "standard"), both have been successfully used to fit galaxy rotation curves. Although the large scattering and measurement errors of the galaxy rotation data is compatible with a wider range of $\alpha$ values, it is important to emphasize that the parameter $\alpha$ is solely determined from the terrestrial *G*-analysis, and not by a fit to the galaxy rotation data.

Within a recent comprehensive review about *G* experiments the authors concluded with the remark "The situation is disturbing—clearly either some strange influence is affecting most G measurements or, probably more likely, the measurements have unrecognized large systematic errors" [24]. The presented analysis describes a possible scenario for "some strange influence", as a serious alternative to "unrecognized large systematic errors". On the other hand, among the *G* experiments being published to date, the selection of Quinn's and Schlamminger's data is subjective, since other experiments show trends which are not compatible with the model presented here [1,11,14]. In particular, one of the more recent Cavendish-type experiment by Gundlach et al. [11], which operates in a similar acceleration range than the one by Quinn and Speake, has revealed a value close to that reported by Schlamminger et al. However, there is one important peculiarity in Gundlach's experiment: the field masses rotate at the extremely low rate of 20mrad /second on a radius of about 17 cm. In spite of this very slow rate, the corresponding centripetal acceleration is $a_c$=6.8·$10^{-5}$ m/$s^2$, which is 500,000 times larger than $a_0$. Therefore, the reduced mass is accelerated by a similar amount of relative acceleration and MOND effects may become unmeasurable (smaller than ca. 1 ppm) with the given accuracy of the experiment.

In case of the torsion experiment by Quinn and Speake the pendulum motion is either supressed (in cases of balancing the torque from the gravitational forces by an electrostatic torque, dubbed "servo method") or the amplitude of the oscillation is not bigger than the pendulum motion generated by the gravitational acceleration of the test mass (Cavendish method) [10]. Therefore, this experiment may be unique in terms of the visibility of MOND effects. The time-of-swing method at HUST (referred as HUST-09 [25]), which has resulted in a *G* value in agreement with the CODATA value, may not allow to observe MOND effects either due to a larger amplitude of motion (not quoted in [23]) or due to the fact that the data acquisition time extends over a large number of pendulum periods of about 530 s each (in contrast to 120 s in case of the Quinn-Speake experiment): during the time the pendulum needs to change its equilibrium position upon moving the source masses, the gravity acceleration component of moon and sun which points into the direction of the pendulum motion changes by an amount which is significantly larger than $a_0$. (for example, the acceleration of the moon may change by up to 1800 times $a_0$ over a time span of 500 s, depending on the position of the moon, and the new equilibrium position of the pendulum may stabilize after several pendulum oscillations). Although acceleration by moon and sun do not result in any measurable torque, they create an accelerated frame of reference (with respect to the direction of test mass motion), which may limit MOND effects to an unmeasurable magnitude.

For a linear pendulum, the decoupling of seismic noise is much less efficient than for a torsion pendulum. In case of a pair of linear pendulums [12-14], the micro-seismic-induced motion of each individual pendulum is usually much larger than the distance of the two pendulum bodies, which is measured by a microwave cavity [12,13] or by a laser interferometer [14]. The MOND correction is determined the acceleration of each pendulum, therefore MOND effects are likely to be supressed.

Finally, the extrapolation function favoured by the presented analysis is consistent with recent planetary observations. As shown by Hees et al., MOND theories based on the modified Poisson equation can be ruled out from recent Cassini data on the orbital dynamics in our solar system for the range of extrapolation function being discussed here [26]. However, this analysis only refers to the specific effects of the modified Poisson equation and not to the modification of inertia at low magnitudes.

A critical test which may allow to distinguish between the inertia and field interpretation of MOND theory is the velocity dispersion of the recently discovered Milky Way satellite galaxy CRATER II [27]. As pointed out by Mc Gaugh [21] the MOND prediction for the velocity dispersion with and without considering the external field effect, which does not allow the observation of MOND effects in terrestrial experiment in case of the field interpretation of MOND, differs by a factor of two. Hence, CRATER II is not only a critical test laboratory for MOND vs dark matter, but it should also allow to distinguish between these two interpretations of MOND. For the inertia case the external field effect is irrelevant and the velocity dispersion should be around 4 km/s [21]. The experimental confirmation of this prediction would be a strongly supportive argument for the discussed interpretation of $G$ discrepancies by MOND theory.

**Conclusion**

The presented analysis has revealed a first indication that the observed discrepancies between $G$-values determined by different terrestrial experiments may have the same physical origin as the anomalies of galaxy rotation curves. Since unknown or underestimated systematic errors in current $G$ experiment cannot be ruled out, this finding may just be a coincidence. In order to test this hypothesis, accurate measurements of small forces (gravitational and non-gravitational) in different regimes of acceleration by different methods are pivotal. It is recommended to run Cavendish experiments using a variety of torque magnitudes– either by comparison of different source masses or by using different angular positions for a given source mass array. Measurements of torque ratios may be less sensitive to some systematic errors than absolute $G$ measurements. The dynamics of the experimental procedure (magnitude of pendulum oscillation, pendulum period, source mass movement, position of moon and sun during data acquisition) may have an influence on the results due to the subtle non-linear nature of MOND effects. Linear pendulum experiments seem to be less suited than Cavendish experiments, because they are more prone to strong and uncontrolled pendulum motions driven by micro-seismic activities. The presented model provides a tool and a guideline for data analysis within the inertia interpretation of MOND theory.

**Acknowledgements:**

The author likes to express his thanks to Clive Speake, University of Birmingham, UK, for checking the consistency of the analysis of his $G$ experiment, Stacy Mc Gaugh from Case Western Reserve University for providing the galaxy data and Hinrich Meyer, Carsten Niebuhr and Eberhard Wuensch, DESY, Germany, for motivating discussions.